\documentclass[11pt]{article}
\usepackage{moriond,epsfig}

\def\Journal#1#2#3#4{{#1} {\bf #2}, #3 (#4)}

\def\APJ{\em ApJ}
\def\APJL{\em ApJ Letters}
\def\AA{\em A\&A}
\def\AAS{\em A\&AS}
\def\MNRAS{\em MNRAS}
\def\ARAA{\em ARAA}
\def\NAT{\em Nature}

\def\be{\begin{equation}}
\def\ee{\end{equation}}
\def\bea{\begin{eqnarray}}
\def\eea{\end{eqnarray}}

\begin{document}

\title{A NUMERICAL STUDY OF COSMIC SHEAR STATISTICS}

\author{A. Thion$^{1}$, Y. Mellier$^{1,2}$, F. Bernardeau$^{3}$, E. Bertin$^{1,2}$, T. Erben$^{4}$, L. van Waerbeke$^{5}$}

\address{$^{1}$ Institut d'Astrophysique de Paris, 98 bis bvd Arago 75014 Paris, France \\ $^{2}$ DEMIRM, Observatoire de Paris, 61 avenue de l'Observatoire 75014 Paris, France \\ $^{3}$ SPhT, CEA/Saclay, F-91191 Gif-sur-Yvette cedex, France\\ $^{4}$ Max-Planck-Institut f\"ur Astrophysik, Karl-Schwarzschild-Str. 1, 85741 Garching, Germany\\ $^{5}$ CITA, 60 St. George Street, Toronto M5S 3H8, Canada}

\maketitle \abstract{\noindent We explore the stability of the variance and skewness of the cosmic gravitational convergence field, using two different approaches: first we simulate a whole MEGACAM survey (100 sq. degrees). The reconstructed mass map, obtained from a shear map recovered by the usual method via a maximum likelihood method, shows that the actual state-of-the-art data analysis methods can accurately measure weak-lensing statistics at angular scales ranging from 2.5' to 25'. We looked also at the influence of a varying signal-to-noise ratio over the shear map (due to local variations of source density) on the mass reconstruction, by means of Monte-Carlo simulation. The detectable effect at small scales can easily be corrected-for in most of the relevant cases. These results enhance the confidence in the capability of future large surveys to measure accurately cosmologically interesting quantities.}

\section{Introduction}

\noindent The cosmic shear effect has recently been detected (see Bacon et al. 2000, Wittman et al. 2000, Kaiser et al. 2000, van Waerbeke et al. 2000 and these proceedings) and the shear variance has been measured in a range of scales from 30'' to 10'. However, theoretical investigation of lensing statistics (Bernardeau et al 1997) show that the measurement of the third order moment of the convergence $\kappa$ is necessary to break the degeneracy between $\Omega_{m}$ and $\sigma_{8}$ (for symmetry reasons, the third order moment of the shear is null). This implies to proceed to a full reconstruction of the convergence map.
\\
\noindent Knowing that the skewness of the convergence field is a statistic much more difficult to measure than the variance, this raises a few questions: (1) do the measurement errors preserve the correlation between cells? (2) How the measurement errors of the shear propagate into the convergence reconstruction? (3) How to correctly estimate the noise due to the intrinsic ellipticity of galaxies in an observational context? (4) How clustering of the galaxies (leading to a spatially varying S/N ratio) affects this reconstruction?
\\
\noindent The points 1) and 2) have been adressed by van Waerbeke et al. (1999), considering only the noise coming from intrinsic ellipticity. We tried a different approach: we performed two kinds of simulations to partly address all those problems.

\section{The simulated lensing survey}

\noindent We performed the simulation of a large (100 deg$^{2}$) compact survey whose characteristics are similar to those of the future MEGACAM dedicated survey at CFHT. The simulation is generated in 6 steps:\\
First we have created the synthetic catalog of the whole survey (in an open CDM Universe), using the {\em SkyStuff $^{a}$} program. We then modified the relevant galaxy parameters (axis ratio, luminosity, etc.) accordingly to 2-dimensional simulations of dark matter with an angular resolution of 2.5' (see van Waerbeke et al (1999) for details). We generated in this manner numerous sub-catalogs (40$\times$40) and used the {\em SkyMaker} \footnote{SkyStuff and SkyMaker are freely available at {\sf ftp://ftp.iap.fr/pub/from\_users/bertin}}  software to create the synthetic images of the galaxies lensed by this 2D simulation. These images have been analyzed with {\em IMCAT} \footnote{ see {\sf http://www.ifa.hawaii.edu/faculty/kaiser/}}, and we used SExtractor to clean the catalogs of spurious detections (and to select the objets with uncorrupted photometry as indicated by the FLAG keyword). The measured shear was then used to reconstruct the $\kappa$ map.

\noindent The characteristics of the survey are listed in table \ref{tab:table1}. As we have only 2-dimensional dark matter simulations for a 10$\times$10 degrees map, all sources in a given area are affected the same way, regardless of their redshift. Considering the wide variety of shear values and galaxy parameters, this should not weaken our conclusions. The shear is estimated by the quantities $\frac{<e_{1}>}{<P^{\gamma}_{11}>}$ and $\frac{<e_{2}>}{<P^{\gamma}_{22}>}$, where the $P^{\gamma}$'s are components of the shear susceptibility tensor and $<>$ means a {\em spatial} average.

\begin{table}[t]
\begin{footnotesize}
\vspace{0.1cm}
\begin{center}
\begin{tabular}{|c|c|}
\hline
Size & 100 sq. degrees \\
\hline
Catalog initial density & $\approx$ 31 galaxies/arcmin$^{2}$\\
\hline Density of detected objects & $\approx$ 22 galaxies/arcmin$^{2}$\\
\hline
Catalog features & $m_{gals} \in [17,24.5]$, LMC ext. law \\
\hline
Image pixel size & 0.2'' \\
\hline
Seeing FWHM & 0.7'' \\
\hline
PSF anisotropy & 0.05'' rms/axis (tracking)\\
\hline
Sky SB, read-out noise & 20 mag/arcsec$^{2}$ (I Band), 5e$^{-}$/ADU \\
\hline
Object selection & FLAG = 0, $|e_{1,2}| \leq 1$ \\
\hline
Stars & $m_{*} \in [10,21.5]$, 60 usables/image\\
\hline
\end{tabular}
\end{center}
\caption{\tiny{Characteristics of the simulated survey: the catalog is obtained from a Schechter function and a PDE prescription, and designed to reproduce observations. The images are simulation of CCD frames taken at CFHT. \label{tab:table1}}}
\end{footnotesize}
\end{table}

\noindent To correct for the noise coming from intrinsic ellipticity, 
 we have simply calculated 
$\frac{<e^{2}_1>}{<P^{{\gamma}^{2}}_{11}>}$. 
Since the theoretical noise is defined from the complex ellipticity,\\
\be
\epsilon =  \frac{a-b}{a+b} \,  e^{i2 \theta}
\label{eq:eps}
\ee \\

\noindent which is not equal to the KSB-defined ellipticity (which is itself filtered by Gaussians windows, etc.) it is interesting to get such a simple estimator. Nevertheless, this estimation is in very good agreement with the noise one can measure directly on the power spectrum of $|\gamma|$, indicating that we obtain a white noise. We applied the results of Bernardeau et al (1997) to correct for the cosmic variance, using a `sample size' equal to 6 times our largest scale of measure, i.e. 2.5 degree. The variance and skewness of $\kappa$ that we obtain as a function of the filtering scale are shown in fig 1. \\

\noindent Although the simulated images still have a quality superior of those of real data (e.g. no strong PSF anisotropy - we still performed the anisotropy correction -, no big diffraction spikes, no bad columns, no strong gradients), the accuracy of those measures shows that current measurement techniques, applied to high-quality data, are reliable enough to allow a good determination of large-scale lensing statistics.

\section{Varying the S/N ratio on the shear map}

In the weak lensing approximation one usually estimates the shear inside a window by the mean of the galaxies' ellipticities inside the window, i.e.\\
\be
\tilde{\gamma}_{\alpha} = \gamma_{\alpha} + \frac{\sum^{N}_1 \epsilon_{\alpha}}{N}
\label{eq:estim1}
\ee \\
\noindent So one considers the variance of the estimator\\
\be
<\tilde{\gamma}^{2}>_{\epsilon,N} \simeq <\gamma^{2}> + \frac{\sigma_{\epsilon}^{2}}{\bar{n}}
\label{eq:var1}
\ee \\

\noindent where $\bar{n}$ is the mean number of objects per window and $<>_{\epsilon,N}$ means averaging on the ellipticity distribution and the spatial distribution of tracers. It is then clear that the signal-to-noise ratio of the shear depends on the local density of tracers, leading to local noise variation. To investigate this effect, we performed a simulation of a 25 deg$^{2}$ survey, again with 2-dimensional shear maps; the positions of tracers were drawn either from a random (Poisson) distribution, or as the result of a random-walk process, of which third and fourth moment are independant of scale, while the angular autocorrelation function of the galaxies has an amplitude roughly one order of magnitude higher than that of faint blue galaxies ($\sim$ 0.15 at 1', see e.g. Brainerd \& Smail (1998)). We used 60 realizations of the dark matter maps, and for each of them we have distributed intrinsically elliptical tracers ($\sigma_{|\epsilon|} \approx 0.2$), distributed in the two differents manners, and we studied 3 different densities (10, 30 and 200 objects/arcmin$^{2}$).

\begin{center}
\begin{figure*}[t]
\centerline{\hbox{
\psfig{figure=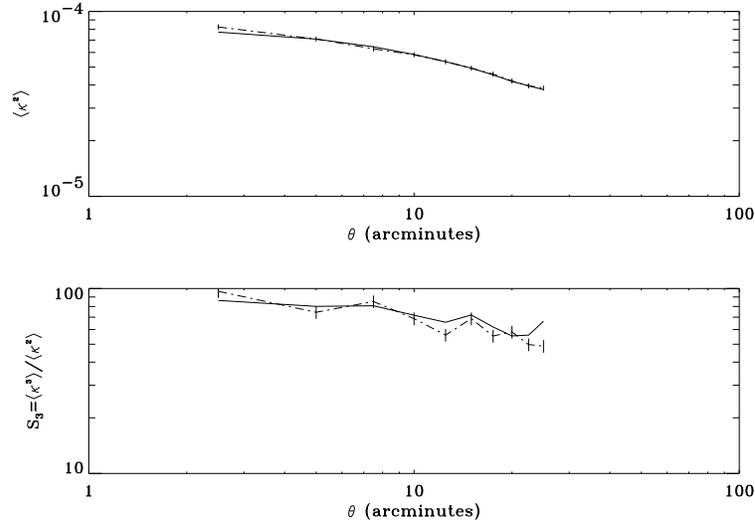,bbllx=8pt,bblly=7pt,bburx=500pt,bbury=
350pt,height=7cm}
}}
\caption{\tiny{Variance and skewness of the convergence field measured from the simulated survey (dash-dotted lines) compared to the theoretical values (solid lines). The ``error'' bars are the minimal errors calculated by van Waerbeke et al (1999). }}
\end{figure*}
\end{center}

\noindent The results on the variance for $ \bar{n} $ = 30 galaxies/arcmin$^{2}$ are show in fig. 2.; an excess of power at small scales is visible in the case of clustering of galaxies. Note however that in this set of simulations galaxy positions are uncorrelated with dark matter distribution (Bernardeau 1998). Although this is not presented on a plot, we checked that the effect on the skewness is only due to the alteration of the variance. To correct for the effect, one has to realize the simple fact that the noise correction must not be done by means of eq. (\ref{eq:var1}), but instead that\\
\be
<\tilde{\gamma}^{2}>_{\epsilon,N} \simeq <\gamma^{2}> + \sigma_{\epsilon}^{2}\times<\frac{1}{N}>
\label{eq:var2}
\ee \\

\noindent where $<\frac{1}{N}>$ is filtered at the scale of interest. For strongly correlated, low-density samples, the $<\frac{1}{N}>$ statistic can significantly differ from $\frac{1}{\bar{n}}$. We can see on fig. 2. that this suffices to solve the problem. Such a correction is not as accurate in the $\bar{n}$ = 10 galaxies/arcmin$^{2}$ case, but this should not constitute a problem in an observational context as the current surveys have galaxies samples with larger density and correlation length. 

\begin{center}
\begin{figure*}[thbp]
\centerline{\hbox{
\psfig{figure=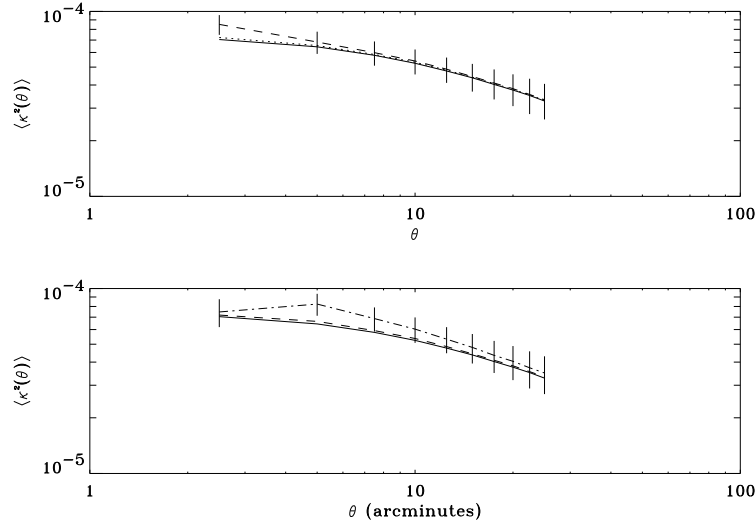,bbllx=8pt,bblly=7pt,bburx=500pt,bbury=
350pt,height=7cm}
}}
\caption{\tiny{The variance of the convergence field for a 25 sq degrees survey. {\em Top panel} \ : the tracers density is 30 objects/arcmin$^{2}$ and a the ellipticity noise correction made by $\frac{\sigma_{\epsilon}^2}{\bar{n}}$. The solid line is the theoretical value, the dotted line is the value obtained for a poisson distribution of tracers, the dashed-line is obtained when the tracers positions are drawn from a random-walk. {\em Bottom panel}\ : the dashed line is the variance of the random-walk case, with noise corrected as $\sigma_{\epsilon}^2\times<\frac{1}{N}>$. The dash-dotted line is the same case with a tracer density of 10 objects/arcmin$^{2}$. For clarity only error bars of the `problematic' cases are shown.}}
\end{figure*}
\end{center}

\section{Conclusion}

\noindent In the framework of future wide field surveys (e.g. the MEGACAM lensing survey), it it essential to verify the stability of the statistics of the density field through the measurement process. We checked this stability in two ways: first we performed a detailed simulation of a large (100 deg$^{2}$), compact survey and analyzed the obtained images, and we succeeded to extract the cosmological signal. Secondly, we investigated the effect of rapidly varying S/N ratio on the convergence reconstruction via realizations of a random-walk process; the effect introduces spurious variance at small scales but can easily be corrected for in most relevant cases. This work therefore enhances our confidence on the capability of future wide surveys to accurately measure cosmic shear statistics.

\section*{Acknowledgments}
This work was supported by the TMR Network ``Gravitational Lensing: New Constraints on Cosmology and the Distribution of Dark Matter'' of the EC under contract No. ERBFMRX-CT97-0172. A. T. is grateful to the MAGIQUE center (IAP) for providing storage and computational facilities. The authors would also like to thank the organizers of the Rencontres de Moriond.


\section*{References}
\begin{thebibliography}{99}


\bibitem{bacon} Bacon, D. J., R\'efr\'egier, A., Ellis, R. S., \Journal{\MNRAS}{}{}{2000}, submitted
\bibitem{bart96}Bartelmann, M., Narayan, R., Seitz, S., Schneider, P., \Journal{\APJ}{464}{L115-L118}{1996}
\bibitem{barsch00} Bartelmann, M., Schneider, P., astro-ph/9912508, submitted to Elsevier Preprint
\bibitem{bern98} Bernardeau, F., \Journal{\AA}{338}{375:382}{1998}
\bibitem{bern97}Bernardeau, F., van Waerbeke, L., Mellier, Y., \Journal{\AA}{322}{1:18}{1997}
\bibitem{sex}Bertin, E., Arnouts., S., \Journal{\AAS}{117}{393-404}{1996}
\bibitem{hh}Hoekstra, H., Franx, M., Kuijken, K., Squires, G., \Journal{\APJ}{504}{636}{1998}
\bibitem{brain}Brainerd, T. G., Smail, I., \Journal{\APJ}{494}{L137B}{1998}
\bibitem{ksb95} Kaiser, N, Squires, G. \& Broadhurst, T., \Journal{\APJ}{449}{460}{95}
\bibitem{kwl00} Kaiser, N., Wilson, G., Luppino, G., astro-ph/0003338, \Journal{\APJL}{}{}{2000}, submitted
\bibitem{lk97} Luppino, G. A., Kaiser, N., \Journal{\APJ}{475}{20}{1997}
\bibitem{mell99}Mellier, Y., \Journal{\ARAA}{37}{127:189}{1999}
\bibitem{vwbm99}van Waerbeke, L., Bernardeau, F., Mellier, Y., \Journal{\AA}{342}{15:33}{1999}

\bibitem{vW00} van Waerbeke, L. {\it et al.}, \Journal{\AA}{}{}{2000}, in press

\bibitem{wittman}Wittman, D. M. {\it et al.}, \Journal{\NAT}{405}{143-148}{2000} 
\end{thebibliography}
\end{document}